# Institutional Collaboration Recommendation: An expertise-based framework using NLP and Network Analysis

**Authors**

**Hiran H Lathabai, Ph.D.**
Department of Computer Science,
Banaras Hindu University,
Varanasi-221005 (India).
Email: hiranhl007@gmail.com

**Abhirup Nandy, M.Sc.**
Department of Computer Science,
Banaras Hindu University,
Varanasi-221005 (India).
Email: abhirupnandy.online@gmail.com

**Vivek Kumar Singh, Ph.D.**
Department of Computer Science,
Banaras Hindu University,
Varanasi-221005 (India).
Email: vivek@bhu.ac.in

**Corresponding Author:**

**Prof. Vivek Kumar Singh**
Professor & Head,
Department of Computer Science,
Banaras Hindu University,
Varanasi-221005 (India).
Email: vivek@bhu.ac.in

# Institutional Collaboration Recommendation: An expertise-based framework using NLP and Network Analysis

**Abstract:** The shift from 'trust-based funding' to 'performance-based funding' is one of the factors that has forced institutions to strive for continuous improvement of performance. Several studies have established the importance of collaboration in enhancing the performance of paired institutions. However, identification of suitable institutions for collaboration is sometimes difficult and therefore institutional collaboration recommendation systems can be vital. Currently, there are no well-developed institutional collaboration recommendation systems. In order to bridge this gap, we design a framework that recognizes thematic strengths and core competencies of institutions, which can in turn be used for collaboration recommendations. The framework, based on NLP and network analysis techniques, is capable of determining the strengths of an institution in different thematic areas within a field and thereby determining the core competency and potential core competency areas of that institution. It makes use of recently proposed expertise indices such as *x* and *x(g)* indices for determination of core and potential core competency areas and can toss two kinds of recommendations: (i) for enhancement of strength of strong areas or core competency areas of an institution and (ii) for complementing the potentially strong areas or potential core competency areas of an institution. A major advantage of the system is that it can help to determine and improve the research portfolio of an institution within a field through suitable collaboration, which may lead to the overall improvement of the performance of the institution in that field. The framework is demonstrated by analyzing the performance of 195 Indian institutions in the field of 'Computer Science'. Upon validation using standard metrics for novelty, coverage and diversity of recommendation systems, the framework is found to be of sufficient coverage and capable of tossing novel and diverse recommendations. The article thus presents an institutional collaboration recommendation system which can be used by institutions to identify potential collaborators.

1. Introduction

Institutional organization, being one of the three major organizations of science (the other two are intellectual and social organizations) has played a pivotal role in the progress of science. Funding is one of the major fuels of R&D activities. An early focus of funding based on 'trust' outlined the importance of the role of institutions in scientific progress. However, in the past few decades, a shift from 'trust-based' funding to 'performance-based' funding (Sörlin, 2007) forced funding agencies to adopt sharp performance assessment methods. At the same time, this shift also forced institutions to strive for continuous improvement of performance. The rise of many international ranking systems such as QS, THE, ARWU, CWTS, etc., is a natural consequence of the shift to 'performance-based funding' and several funding agencies rely on these rankings. Some funding agencies prefer 'thrust area performance' as a yardstick for fund allocation. National agencies in many countries are entrusted to formulate national strategies for nurturing institutions of excellence in thrust/priority areas. For instance, back in 2006, an Indian working group on thrust areas hand-picked cyber security, multi-scale modelling, Quantum theory and applications, etc., as some of the thrust areas in engineering sciences. The establishment of the Interdisciplinary Cyber Physical Systems (ICPS) division by the Department of Science and Technology (DST), Govt. of India is another evidence for the increasing emphasis on 'thrust area performance-based' funding. When it comes to 'performance-based funding' or 'thrust area performance-based funding', institutions are always required to remain innovative and relevant.

In the case of an institution, its scholarly contributions might span over many fields. A research field comprises of many thematic areas. For example, for the subject 'Computer Science', one can consider 'Software Engineering', 'Data Science', 'NLP', 'Data Mining' etc. as some important thematic areas. The level of contribution of an institution with respect to a thematic area in a given field might vary, or in other words the institution may have varied 'thematic research strengths'. An institution can be strong in some areas and relatively less strong in some other areas, and may be even weak in several other areas. For example, an institution may have strong research capability in 'artificial intelligence' but may be relatively weak in 'software engineering', both of which are thematic areas of Computer Science. For being innovative and relevant for 'performance-based funding' and 'thrust area performance-based funding', the institutions must take the responsibility to work towards (i) improving their overall performance (ii) improving their performance in their strong/core competency areas and (iii) identification and conversion of their relatively less strong (or potential competency areas) to core competency areas. Collaboration is one of the effective ways to achieve these goals as collaboration is found to be beneficial to the academic productivity (Lee & Bozeman, 2005; Ductor, 2015; Parish et al., 2018). The rise of team science and research networks is evident since last two decds (Adams, 2012).

Academic collaboration happens within and across nations at different levels- author level and institutional level. Author level collaborations and its benefits are well-explored at different perspectives. Strong pragmatism and a high degree of self-organization were identified as major characterization of author collaborations (Melin, 1996; Wagner & Leydesdorff, 2005). Among the many perspectives of author collaboration studies, research on co-authorship using the network approach was one of the effective approaches. Co-authorship networks are much effective (Katz & Martin, 1997) in the investigation of patterns of scientific collaboration and its dynamics (Newman, 2004), patterns of behavior of collaborating scientists (Martin et al., 2013), the relationship between patterns of collaboration and individual productivity (Parish et al., 2018), etc. Co-authorship networks are utilized for the creation of academic researcher collaboration recommendation systems with the aid of various machine learning methods (Brandão et al., 2013; Meghanathan, 2016; Pujari & Kanawati, 2015).

Unlike author collaboration, institutional collaboration is more difficult to be established, manage, and is relatively more difficult to be successful. Activities associated with an agreement signed in 1988 to promote academic collaborations between UK and Brazil can be regarded as an example of international collaboration at the level of institutions (Canto & Hannah, 2001). A study on the Italian academic research system that investigated the correlation between scientific productivity and institutional collaboration intensity (of international as well as private institutions) failed to observe the correlation between extramural research collaboration and the overall performance of research institutions (Abramo et al., 2009). This affirms the importance of the identification of suitable partners for collaboration and that recommendation systems are always prone to this risk when it comes to institutional collaborations. Recently established ICPS division by DST, Govt. of India to foster a healthy innovation ecosystem to make India a leading player in CPS, that functions on the basis of the hub and spoke model for facilitating productive collaborations between institutions that belongs to academia and industry, can be viewed as a diligent attempt to achieve maximum benefit from collaborative research by overcoming this risk.

Though there are several recommendation systems in varying domains (see for example Pazzani & Billsus, 2007; Lops, De Gemmis, & Semeraro, 2011; Bouraga et al., 2014; Kim & Chen, 2015; Son & Kim, 2017; Shao, Li & Bian, 2021), the academic domain is relatively less explored for recommendations (the 'Related Work' section briefly surveys some of the recent studies). In the academic domain, the particular problem of institutional collaboration recommendation is virtually unexplored. Being effective for collaboration studies, research based on the network approach is found in the related literature. Major studies based on co-institutional networks were oriented around the mapping of institutions including the most collaborative institutions or the study of evolutionary dynamics of institutional collaboration within a field/subfield (Koseoglu, 2016; Deng et al., 2019; Ashraf et al., 2020; Payumo et al., 2021). Usage of co-institutional networks for cross-institutional mapping of two fields is also found (Ye et al., 2012). Despite the effectiveness of the network approach, no attempt to design an institutional collaboration recommendation system is known to us. Inspired by this gap, we propose a framework that can help improve the performance of an institution, so that it can remain relevant for both 'performance based-funding' and 'thrust area performance based-funding'. Our framework is expected to toss recommendations to institutions based on two strategies:

- *Strategy 1* intends to help institutions to enhance/maximize their strength in their core competency areas by suggesting other institutions that also have a core competency in those areas.
- *Strategy 2* intends to help institutions to improve their strength in their potential core competency areas by (i) suggesting high priority institutions that are core competent in those areas and (ii) suggesting low priority institutions that are potentially competent in those areas but are more competent than the institution of our interest in those areas.

The terms high priority and low priority are used because collaborations with high priority institutions (if it happens to be) might enhance the strengths in potential core competency areas at a relatively higher pace than collaboration with low priority institutions. These two strategies will directly help to enhance the thematic research strengths of an institution. If executed successfully, the overall improvement of performance will also be achieved to a great extent.

Our framework, which is an extension of the preliminary framework designed by Lathabai et al. (2021), consists of two sections- the *expertise determination section* and the *recommendation retrieval section*. Expertise determination section functions by determining the strengths of each institution in different thematic areas. Mapping of published works into thematic areas and determination of thematic strengths of each institution is achieved with the help of a state-of-the-art NLP module and the network-based framework for contextual productivity assessment (Lathabai et al., 2017). Core competency thematic areas or areas of expertise of an institution can be determined based on thematic strengths of institutions using expertise indicators such as $x$-index and $x(g)$-index, that are designed according to the principles of $h$-index and $g$-index. The key idea behind our framework is that the top $x$ thematic areas found in the $x$-core (represented by $x$-index) can be treated as the core competency areas and the top areas that occupy positions from $x+1$ to $x(g)$ out of the $x(g)$ areas found in the $x(g)$-core can be treated as potential core competency areas. The job of the recommendation retrieval section is to execute strategies 1 and 2 using the information of core competent and potential core competent institutions in the area and toss recommendations. A detailed description of the framework is given in the Section 3.

Our recommendation system framework is demonstrated using the case study of 195 Indian institutions in the field 'Computer Science'. The framework is validated using major metrics related to novelty, coverage and diversity of recommendations. One of the popular metrics namely, the accuracy of the framework cannot be determined at this stage. It can be verified post the implementation of the system. Upon verification, our framework is found to be (i) effective in terms of coverage and (ii) excellent in tossing suitable recommendations that are novel and diverse. Definition of standard metrics for validation (identified from relevant related literature) and validation exercise of the recommendation system framework are given in section 6. The underlying approach behind our framework is novel and, to the best of our knowledge, there does not exist any institutional collaboration recommendation system that used this approach. Therefore, it is not possible to compare our framework with existing frameworks for collaboration recommendations in academia and industry.

Our system is based on the rationale- 'the enhancement of an institution can be achieved if it is committed to (i) maximize its strength in its strong or core competency areas and (ii) strive towards the conversion of potential core competency areas to core competency areas by strengthening performance in those areas'. Though we do not attempt to develop any new theories or improve existing ones, our work attempts a diligent and well-engineered sequence of application of relevant methods that are strongly rooted in theories from the fields of 'Network Science' and 'Natural Language Processing'. By utilizing the unexplored potential of these fields for the cause of institutional collaboration recommendation, our attempt can be viewed as an effort to bridge the gap between theory and practice. As our system can be useful for improving the performance of research institutions, it can indirectly contribute to the potential scientific advancement and its associated social impact. Furthermore, this may strengthen the importance of the role of the institutional organization of science in scientific progress.

## 2. Related Work

There are several recommendation systems in varying domains (see for example Pazzani & Billsus, 2007; Lops, De Gemmis, & Semeraro, 2011; Bouraga et al., 2014; Kim & Chen, 2015; Son & Kim, 2017; Shao, Li & Bian, 2021). However, recommendations in academic domain are relatively less explored. In academic domain, the previous studies have mainly focused on scientific article recommendation (aka paper recommendation), expert recommendation, publication venue recommendation and recommendations for collaboration between research teams etc. As mentioned earlier, collaboration is a major factor that can improve the quality of research at different levels- individual as well as institutional. For such collaborations to take place, an understanding of the suitable collaborating partners is needed, which in turn can be facilitated by an appropriately designed recommendation system. These collaborations, when they happen, are known to bring in progressive knowledge transfer, resource sharing etc for advancement of knowledge and research. Here we survey some of the relevant such studies on recommendations in academic domain.

Scientific article recommendation is a problem where the system recommends a scientific article to the user based on certain parameters like information need, researcher interest etc. Author-based data has been used for scientific article recommendation in some previous studies (such as Xia, Liu, Lee, & Cao, 2016). Another study (Waheed et al., 2019) also used author-specific data like the previous engagement of the author with the theme to create a multilevel citation network to find important authors and subsequent recommendations.

Several other studies have used data from scientific articles, such as topical relevance, citation proximity, article tags, etc. for recommendations in a network-based framework (Ali et al., 2020). Bibliographic coupling has also been used to create a co-citation based network, alongside a weighted in-text citation scoring method, to increase suggestions of most commonly cited articles (Habib & Afzal, 2019).

Publication venue recommendation is another interesting academic recommendation task pursued by some previous studies. Publication venue recommendation involves suggesting potential publication sources- both journals and conferences- for a scientific article. A good publication venue recommendation system can utilize the basic article data like the title, abstract and references to provide good quality recommendations (Medvet, Bartoli, & Piccinin, 2014; Rollins et al., 2017). Other studies on publication venue recommendation have used a network-based approach such as exploiting the citation network, along with the title and abstract data, to build similarity score matrices between papers and journals (Alshareef, Alhamid, & El Saddik, 2019). Other kinds of networks like social networks and co-authorship networks have also been used to target publication venues based on the author's historical selections (Luong et al., 2012). The location-based social data merged with topic relevant social data has also been exploited for recommendation (Zahalka, Rudinac, & Worring, 2015). A hybrid bibliographic network, which includes topic modelling and centrality measure, has been used previously to provide good suggestions of publication venues (Pradhan & Pal, 2020). Recently, a deep-learning based framework has also been applied on the article data for publication venue recommendation in an optimized manner (Pradhan, Kumar, & Pal, 2021).

Expert recommendation is another area exolored in the domain of academic recommendations. It involves suggesting individual researchers who have expertise in a given research area/ topic. Some previous studies used semantic similarity for recommending experts, using a spatial mapping relationship (Gao et al., 2019). Another approach used a multi-level search for experts within the community to provide suggestions at different academic levels (Yang et al., 2014). Expert recommendations often involve recommendation for research collaboration. Various studies have used different aspects for a probable collaborator recommendation. A majority of these methods are based on social network analysis. One such study involved a co-authorship network-based for collaborator recommendation (Afolabi, Ayo, & Odetunmibi, 2021). This study used the network properties like centrality measures and network coefficients to provide a measure which is then used for recommendations based on link prediction. Another study used co-authorship network and network embeddings methods to learn semantic information among the researchers and then suggest collaborators (Zhang, 2017). One important factor that has been exploited with the co-authorship network for recommendations is the author's demographic data, including the author's career age or experience, and gender (Sun, Lu, & Cao, 2019). The leading author's data can also be collected and utilized to score articles based on the impact left by the leader in the co-authorship network, i.e., proficiency, trend analysis, etc. (He, Wu, & Zhang, 2021). Topic clustering and trend analysis of the articles has been implemented along with the co-authorship network, to provide a researcher's profile to the recommendation system (Kong et al., 2016, 2017). Co-inventorship in patents' data, with information including link semantics, has also been used to retrieve various insights and provide recommendations (George et al., 2021).

Apart from co-authorship networks, co-citation networks and bibliographic networks have also been used for expert recommendation. A basic bibliographic coupling network, along with data like collaboration frequency, previous collaboration history, etc. was utilized for recommendation (Yang et al., 2015). Another heterogenic bibliographic network-based approach (involving both author data and publication data), along with random walks has been used to provide ranked author recommendations (Zhou et al., 2017). Some previous studies have also exploited Machine Learning model along with Discriminant analysis for industrial R&D collaboration recommendation between different industries (Jun, Yoo, & Hwang, 2021). A hybrid approach using both the co-authorship network and the bibliographic network to connect research projects with new researchers has also been explored (Liu, et al., 2019). Recommendation systems development for inventor collaboration is found to be relatively underexplored as compared to academia. A link semantics and supervised learning-based framework by Wu et al. (2013) is one of the two major attempts for inventor collaboration system framework development. As a response to the need for a recommendation system that is not intensive on the usage of link semantics, George et al. (2021) devised a minimal link semantic (MLS) framework for inventor collaboration recommendation system based on co-inventorship networks to retrieve various insights and provide recommendations.

Although there are several studies for collaboration recommendation at the level of individuals, the institutional collaboration recommendation is relatively less explored. Among the limited such studies, the focus has been mainly intra-institutional collaboration. One such study used a social-network approach to compute collaboration index between two bodies, which are intra-institutional (Parada et al., 2013). Another small study included a curated list of six universities, whose data- international academic and research orientations- were observed and a system was devised to provide recommendations for collaboration among them (Hernandez-Gress, Ceballos, & Galeano, 2018). There are, however, not well-developed institutional collaboration recommendation systems and this makes our work important as it tries to fill this research gap. A institutional collaboration recommendation system can exploit the information about research strengths of institutions and suggest suitable collaboration partners to help improve their research capabilities. Such collaborations can help in complementing the research capabilities of paired institutions and hence result in an overall improvement of research capabilities of both the institutions. The framework proposed in this paper presents a systematic effort towards this end.

3. **Proposed Framework**

Determination of research strengths of an institution in different thematic areas (and thereby determining the core competency and potential core competency areas) is the key task of the framework. Retrieval of suitable recommendations based on previously mentioned diligently designed strategies 1 and 2 is the second major task. Thus, our recommendation system framework has two sections-(i) the *expertise determination section* and (ii) the *recommendation retrieval section*. A schematic diagram of the framework showing both the sections is given in **fig 1**.

Determination of the thematic strengths of an institution is the first step. For that, an institution's scientific publications have to be mapped to respective thematic areas and then metrics that reflect the performance of an institution with respect to thematic areas have to be decided and method(s) to compute these scores are to be designed. The foremost concern regarding this is how to determine the thematic areas of research? As scientific literature can be treated as a body of knowledge and several levels of representation of knowledge is possible for it viz., level of thematic areas/subfields, fields of research, major disciplines or broad subjects, there is no proper way or hard and fast rules to define the confinements of each level.

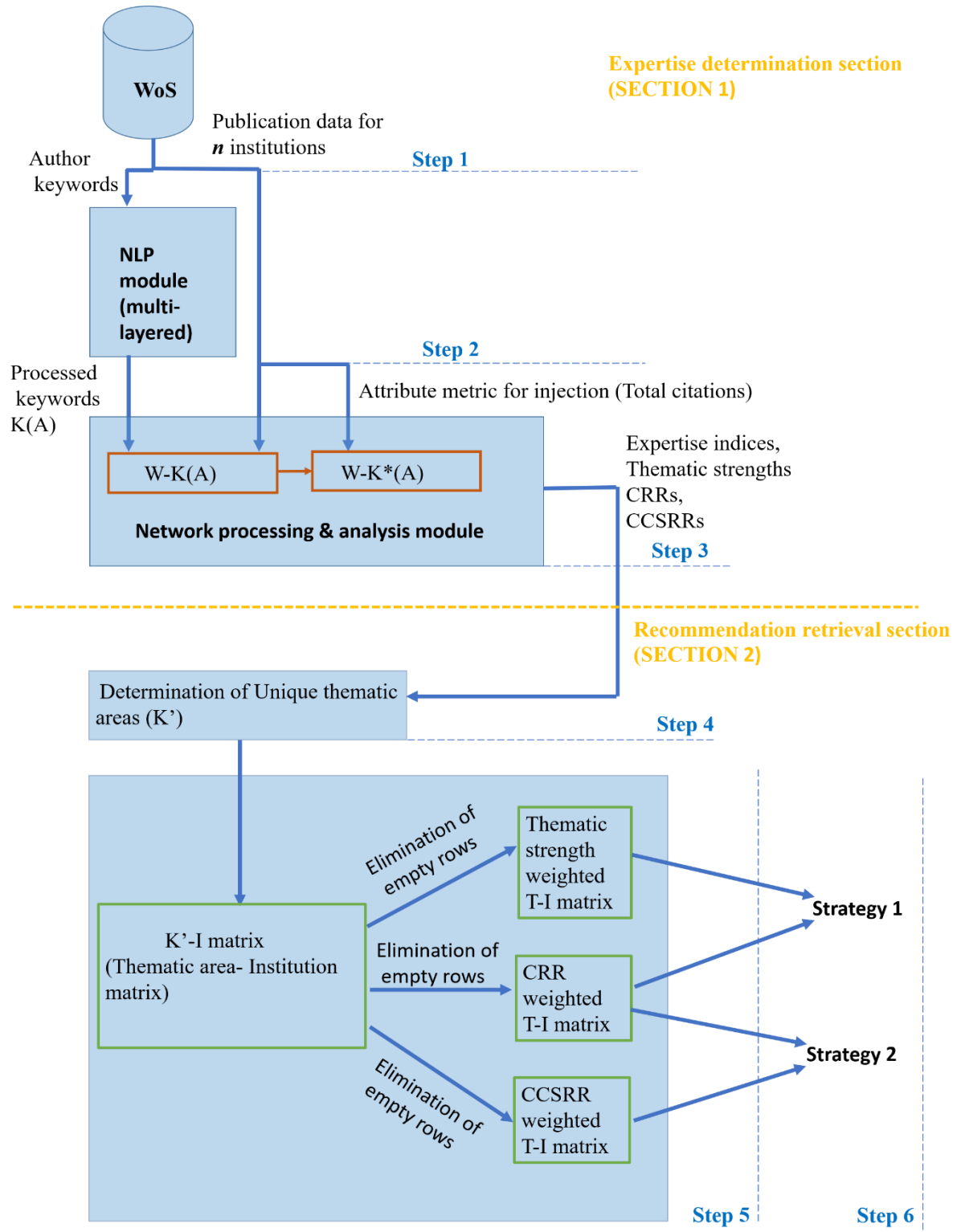

**Figure 1**: Schema of the recommendation system

Indexing systems and databases are designed to permit information retrieval based on certain terms known as 'keywords' and logical combination of such keywords so that accuracy of the retrieval depends on the choice of keywords or logical combination of keywords. Therefore, we can treat keywords as a kind of basic unit that can impart different levels of classification of the scientific literature. A major conjecture that can be formed here is that 'keywords can be used to represent thematic areas or finest level of classification of scientific literature'. A shred of evidence that strengthens this conjecture is the usage of keywords provided by the author in scientific publications as a marker of the specific contribution of an article and the thematic areas within which that contribution can be incorporated. Several text mining studies use many well-developed methods to extract keywords from the article. Databases like Web of Science (WoS) have an internal mechanism to extract keywords and WoS's extracted keywords are known by the name 'Keyword Plus' keywords. Zhang et al. (2016) found that both, the author keywords and the keyword plus keywords are almost equally effective for the investigation of knowledge structure within scientific fields. Therefore, our framework's data processing task starts with the intake of author keywords. Keyword plus keywords and any other extraction algorithm-generated terms can also be used for this purpose. But in this work, we consider only author keywords.

Now the step-by-step procedure of the framework is discussed.

**Step-1: Data collection**

The tasks of Section 1 commence with data collection from reputed online databases like WoS. Institution wise data have to be collected from a country/region with respect to a research field of interest. We intend to demonstrate the framework for the country India for the time period of 2010 to 2019, and provide recommendations for the field 'Computer Science'. More details of the data collection are given in section 4.

**Step-2: Data pre-processing using NLP module**

As we have already mentioned, author keywords, present in documents (articulated in the WoS file as a column with field tag 'DE'), can be used to designate thematic areas, for a start. However, some of the author keywords couldn't be readily used as they might result in loss of accuracy of the analysis since – the terms are too generic, or the terms are too fine-grained for the theme or subject area. For example, terms like 'Data Analysis' or 'Hierarchical Clustering' are two terms that are either too generic or too much fine-grained, to be used as thematic areas of Computer Science. As both of these types of keywords cannot qualify as thematic area designations and there is no hard and fast rule to determine whether these qualify or not, the framework incorporates a natural language processing (NLP) module as an attempt to improve accuracy by reducing the problems that might arise with the raw usage of author keywords. A multi-layered NLP module is used to ensure as much accuracy as possible. We plan to use a Machine Learning model since ML models are easily modifiable to be used in various fields with adequate accuracy in results (Karasu et al., 2018; Sezer & Altan, 2021). Word2Vec is such a neural network model that can perform well in NLP tasks.

*NLP Module*

The raw author keywords might have many ambiguities and may also be prone to the presence of plural and singular versions of some terms. The use of a multi-layered NLP module- one layer for conversion of ambiguous terms to more suitable terms that can represent the thematic area and another layer for replacement of plural words with its singular counterparts, might reduce the size of the raw set (approach illustrated in **fig. 2**). Apart from the improvement of accuracy, this will be useful to reduce computational costs in the recommendation retrieval section.

To improve the accuracy of our model, we used a word embedding model to group together semantically similar keywords under their most popular group member. Word Embedding models are found to be more successful in computing similarity between words (Jatnika et al., 2019; Jurgens et al., 2012). Such word embeddings convert a word into a vector, which can be easily used to find out the similarity between different words. One such widely used model includes "word2vec" (Mikolov et al., 2013). The word2vec model is a shallow neural-network model, which can efficiently take a word as input, and provide $n$ number of similar words as output based on cosine similarity of word vectors. These words can be used to transform our generic and fine-grained words into a usable collection of terms within a thematic area. The model used was trained from scratch, using a dataset of meta-data of scientific research publications. A subset of the Semantic Scholar Open Research Corpus dataset (Ammar et al., 2018) is used to train the model. Only English articles were used in the training, where the raw text from the title and abstract of each article were extracted. This text was then tokenized and fed into the model, which trained on these words. Since we had a very large dataset, we trained the model on 5 epochs, since any more epochs would have increased training time, without a significant increase in accuracy (Caselles-Dupré et al., 2018; Vasile et al., 2016). The model then created a vocabulary specific to the input data. This domain-specific vocabulary is then used to train the model with hopes of achieving better accuracy at scientific terms used in research publications. Word2vec uses these vocabulary terms as its dictionary to store word vectors for each word and find similarities between the words. After training the model, it can be used to infer vectors for a word based on the context in which it has seen the word previously. So we provide the author keywords as input and get a set of keywords as output from the model. This set is limited to top 5 most similar keywords for each author keyword. Then we take a this set of 6 keywords (one author keyword + five derived similar keywords) and try to find the keyword which is most frequent. This is done by storing a row in a table a set of 5 keywords for each author keyword. This author keyword is then replaced at all places with the most frequent one. This provides some level of generalization for very fine grained author keywords. Also, certain generic keywords also fall into their desired category when the rest of the author keywords as used for a document. For example, 'Hierarchical Clustering' and 'Data Analysis' may both be present in a document. After the word2vec step, the most frequent keyword to replace 'Hierarchical Clustering' may be 'Clustering'. Thus, the article will fall under the thematic area of 'Clustering'.

Now, we also see another problem with author keywords. The dataset might have several multi-term keywords where the difference between two keywords may be just their Plural or Singular form. For example, 'Algorithms' and 'Algorithm' may have slightly different cosine similarities with each author keyword, and hence result in ambiguous sets. To solve this problem, a simple lemmatization or Stemming would not achieve the plural to singular transformation as it might incur a loss of data. For example, we may be losing out on data where abbreviations are used as author keywords. The way to solve this problem is to look out for words like 'CNNs' and 'CNN', where the trailing 's' can be removed to convert singular to plural.  Levenshtein distance (Levenshtein, 1966) was used for the second layer to replace plurals with singulars of terms. From each pair of keywords with edit/Levenshtein distance of 1, we have to check whether the last character of the term is 's'. If that is the case, the other term in the pair can be taken as the singular form and every occurrence of the plural term can be replaced with the other term. Any other complex conversion of singular to plural forms (and vice-versa) was handled with the word2vec model where most similar words were used to replace the plural to singular form (since they ought to be used in similar contexts). An example of the usage of a multi-layered NLP module to rephrase keywords is given below.

Several variants of the thematic area 'Cyber physical system' including its plural term were found and with the help of two layers all such occurrences are rephrased as 'Cyber physical system'.

| **Author Keywords** | **NLP processed author keywords** |
|---|---|
| Cyber physical system | Cyber physical system |
| Cyber physical systems | Cyber physical system |
| Cyber physical system (cps) | Cyber physical system |
| Cyber physical system cps | Cyber physical system |
| Cyber physical system-cps | Cyber physical system |

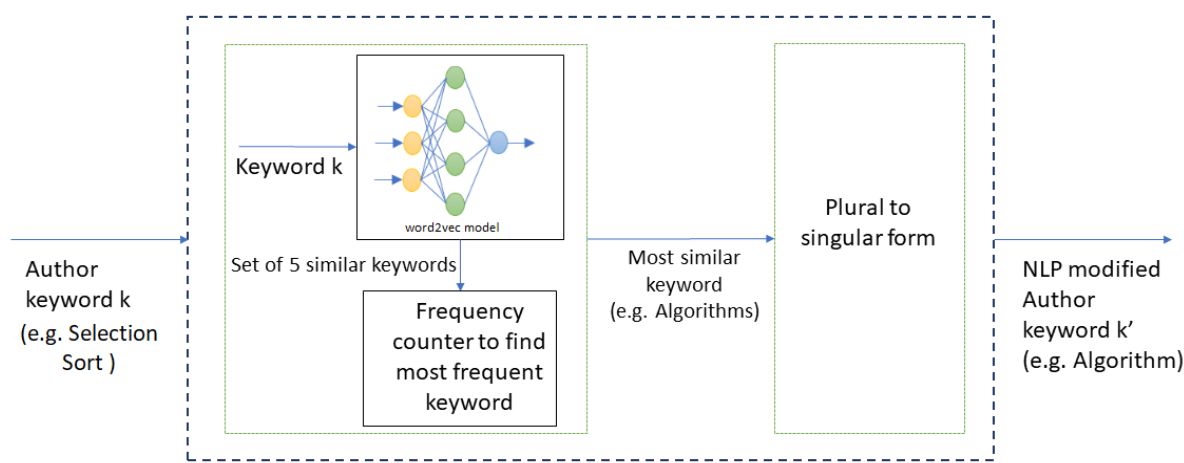

**Figure 2: Multi-layered NLP module**

Let K(A) denote the set of NLP processed author keywords. Now, K(A) represents thematic areas and we can discuss the determination of thematic area strengths of institutions.

**Step-3: Data processing using Network creation and analysis module**

As the raw author keywords are rephrased in the best possible way using state-of-the-art NLP techniques to represent thematic areas, the next major task is mapping of publications of institutions to these thematic areas and computation of suitable scores that reflect the thematic strength. The network-based framework introduced by Lathabai et al. (2017) is capable of this mapping and also serves as a means to compute different kinds of scores with respect to different attributes used as input through Injection methodology. The framework can be used to work upon attributes whose information is not available in the dataset and compute corresponding scores. The full potential of the framework is not utilized here as it is beyond the scope of this work. Here, we use only total citation counts for Injection and computation of thematic strength. The contextual productivity framework by Lathabai et al. (2017) works based on the creation of affiliation networks and weighted affiliation networks. A brief description of these terms is given next.

Networks are structures that consist of vertices that can represent some real-world entity, connections/links between vertices that can represent a relationship between those real-world entities and information/attributes of vertices and/or the links. Networks can be unweighted or weighted depending on the presence or absence of vertex weights or link weights that usually represent the real or relative magnitude of the attributes related to vertices or links. Networks can be unimodal and multi-modal based on the presence of one or more types of vertices/entities. For instance, a citation network of scientific/patent publications is a unimodal network or 1-mode network and an affiliation network of the publication-author network is a 2-mode network (simplest form of multi-modal networks). In the 2-mode publication-author network, the set of publications is the first mode and the set of authors is the second mode. Links will be directed from the first mode to the second mode and there will not be any connections among vertices of the same set/mode as 2-mode networks are essentially bipartite. As we are required to map publications to thematic areas, we have to create a publication-thematic area affiliation network or W-K(A) affiliation network. The formal definition of the W-K(A) network is given below following the notations by Batagelj (2012).

W-K(A) network or the affiliation network of works/publications and author keywords, is a structure W-K(A)= (W, K(A), L, Y) where W is the set of works that forms the first mode and K(A) is the set of keywords that forms the second mode and L is the set of directed links which originate from W and terminate at K(A), Y is the set of the weight of links. W-K(A) affiliation network can be represented by:

$$Y = [y_{wk(a)}]$$

where,

$$y_{wk(a)} = \begin{cases} y(w, k(a)), & if\ w, k(a) \in L \\ 0, & otherwise \end{cases}$$

For unweighted networks, the weights of links will be 1. As we mentioned earlier, the contextual productivity framework creates a weighted affiliation network using relevant attributes through the Injection method. The Injection method facilitates the assignment of weights to links of an affiliation network based on attribute values (computed internally or gathered externally) of the vertices in the first mode. The weighted W-K(A) network will be link weighted using citation scores of the publications and this citation weighted W-K(A) network is denoted as W-K*(A).

*Thematic area strength determination*

With proper choice of attribute that can reflect the strength of an institution in a thematic area, our framework helps to compute thematic area strength using weighted indegree. Before going to that, a very brief definition of degree and weighted indegree are given below:

The degree of a vertex in an unweighted/weighted network is the total number of links associated with it. The weighted degree of a vertex in a weighted network (link weighted network) is the sum of weights of all the links associated with it.

The weighted in-degree analysis of second mode vertices (i.e, author keywords) of the W-K*(A) network will give the thematic area strengths i.e., citations received by an institution with respect to each thematic area via a number of publications in that thematic area. For example, suppose institution *i* has published works w1, w2 and w3 and they received citations 2, 5 and 10 respectively. Let w1 be mapped to thematic areas t1, t2 and w2 be mapped to t1 and t3 and w3 is mapped to t2, t3 and t4. This is shown in **fig 3**. Then thematic area strength of *i* in thematic areas t1 can be obtained as the sum of citations earned by w1 and w2. Therefore,

thematic area strength with respect to t1= 2+5=7

similarly, thematic area strength with respect to t2=2+10=12

thematic area strength with respect to t3=5+10=15

thematic area strength with respect to t4=10

If instead of citations, any other attribute (say altmetric scores from an external database) is used, then also the framework facilitates the injection of these scores and determination of the strength based on that score via the computation of weighted indegree.

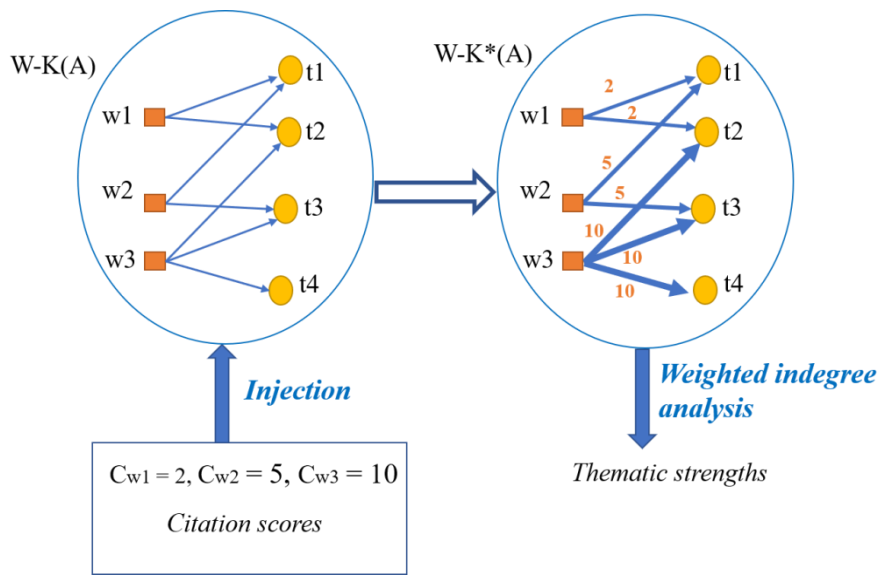

**Figure 3. Illustration of Injection method for thematic strength determination**

*Determination of areas of expertise*

Once the thematic area strengths are computed, we can identify the core competency/expertise areas of an institution as well as the potential core competency/expertise areas. We need to design metrics that can reflect the expertise of institutions based on thematic area strengths and represent it as a single number that reflects the expertise of an institution within that field. These indices can be generally termed as expertise indices. In the bibliometric/scientometric literature, *h*-index (Hirsch, 2005) and *h*-type indices such as *g*-index (Egghe, 2006) and *ψ*-index (Lathabai, 2020), etc., are found. The *h*, *g* and *ψ* indices can partition the publication profile of an author/inventor into productive cores and tails such as *h*-core and *h*-tail, *g*-core and *g*-tail, and *ψ*-core and *ψ*-tail, respectively. Inspired by the *h* and *g* indices, expertise indices namely *x* and *x(g)* indices (Lathabai et al., 2021; Lathabai et al., 2021a) can be defined in the following way:

***x-index***: An institution is supposed to have an *x*-index value of *x* if it has published papers in at least *x* thematic areas and received the score (citation, altmetric score, etc.,) of at least *x* in those areas. These *x* areas that form the *x*-core can be treated as the core competency areas of the institution.

***x(g)-index***: An institution is supposed to have an *x(g)*-index value of *x(g)* if it has published papers in at least *x(g)* thematic areas such that the average score received from these areas amounts at least to *x(g)*. These *x(g)* areas that form the *x(g)*-core can also be treated as the core competency areas of the institution.

We do not attempt the design of an expertise index based on $\psi$-index because it can be computationally more costly and might provide too many recommendations.

*Computation of x-core and x(g)-core from weighted indegree results*

From the sorted profile of an institution (sorted by weighted indegrees of keywords or thematic strengths), the *x*-index of an institution can be computed. *x*-index satisfies the following condition.

$$x = \{\,max_r\,:\,weighted\ indegree\ of\ keyword\ at\ position\ r\ \geq r\,\}$$

*x(g)*-index satisfies the following condition.

$$x(g) = \{\,max_r\,:\,\sum_r weighted\ indegree\ of\ keyword\ at\ position\ r\ \geq r^2\,\}$$

where *r* is an arbitrary rank/position of sorted keywords (sorted by weighted in-degrees).

Expertise index value (value of *x* and *x(g)* indices) for an institution in a field will be different from the expertise index value in another field. This indicates the advantage of using expertise indices for 'thrust area performance-based' funding. Thus, policymakers or funding agencies can identify the deserving institution in thrust areas rather than considering the overall performance indicators like *h*-index, publication and citation counts, etc.

The above-mentioned conditions can also be stated in the following way:

$$x = \{\,max_r\,:\,\frac{weighted\ indegree\ of\ keyword\ at\ position\ r}{r} \geq 1\,\}$$

$$x(g) = \{\,max_r\,:\,\frac{\sum_r weighted\ indegree\ of\ keyword\ at\ position\ r}{r} \geq r\,\}$$

Let $CRR_r = \frac{weighted\ indegree\ of\ keyword\ at\ position\ r}{r}$

$$= \frac{thematic\ strength\ at\ position\ r}{r}$$

be the citations to rank ratio and

$CCSRR_r = \frac{\sum_r weighted\ indegree\ of\ keyword\ at\ position\ r}{r}$

$$= \frac{sum\ of\ thematic\ strengths\ upto\ position\ r}{r}$$

be the cumulative citations to squared rank ratio.

Thus, thematic areas related to institutions that satisfy CRR $\geq 1$ will be the core competency areas. Thematic areas that satisfy CCSRR $\geq 1$ and are not found in the list of core competency areas are the potential core competency areas of an institution. Based on these, strategies 1 and 2 can be designed. Thus, the computation of expertise indices and associated CRR and CCSRR values marks the end of section 1. Now, further steps fall under the second section of the framework.

**Step 4:** *Determination of unique set of keywords/thematic areas*

Let *I* be the set of *n* institutions. For each institution, there will be a set of thematic areas mapped to it via publications. Thematic areas can be mapped to institutions directly for further operations and that requires the identification of a unique set of keywords. Let $k(A)_1$, $k(A)_2$, $k(A)_3$,………..., $k(A)_n$ be the set of thematic areas associated with institutions $i = 1, 2, …………, n$. Let K' be the unique set of keywords, then K' can be determined as:

$$K' = k(A)_1 \cup k(A)_2 \cup k(A)_3 \cup \ldots\ldots\ldots\ldots\ldots k(A)_n$$

Now, institutions can be mapped to thematic areas.

**Step 5:** *Creation of thematic area-institution matrices*

Thematic area-institution matrix K'-I can be created from the institution-specific results. Let it be an $m' \times n$ matrix, where $m' = |K'|$. K'-I matrix is essentially a binary matrix, where 1s indicate an institution have publications (at least one) in the corresponding thematic area and 0s indicate that an institution does not have published in the corresponding area.

Now, we can create weighted matrices from the basic K'-I matrix using information such as thematic strengths (i.e., citations received in each thematic area), CRR values and CCSRR values (of an institution with respect to the thematic area or a thematic area with respect to an institution). When it comes to the thematic strengths of institutions, there can be some thematic areas in which no institutions have strength. This means, though some or all of the institutions might have published works in that thematic area, these works are currently uncited (in future the situation might change) and hence institutions cannot claim any strength in that thematic area. The situation will be different if instead of citations any other metric (say altmetric score) is used. If all the institutions do not have any strength in some thematic areas, all the weighted matrices might have some empty rows (rows with all zero values). These rows should be eliminated from all the three matrices and the resulting matrices will be the useful thematic area-institution matrices that can be designated as thematic strength weighted T-I matrix, CRR weighted T-I matrix and CCSRR weighted T-I matrix, etc. Each of these matrices will be of the size $m \times n$, where $m = |T| \leq m'$. These matrices will be of the form:

$$\text{Citation weighted T} - \text{I} = \begin{pmatrix} C_{11} & \cdots & C_{1n} \\ \vdots & \ddots & \vdots \\ C_{m1} & \cdots & C_{mn} \end{pmatrix}$$

$$\text{CRR weighted T} - \text{I} = \begin{pmatrix} CRR_{11} & \cdots & CRR_{1n} \\ \vdots & \ddots & \vdots \\ CRR_{m1} & \cdots & CRR_{mn} \end{pmatrix}$$

$$\text{CCSRR weighted T} - \text{I} = \begin{pmatrix} CCSRR_{11} & \cdots & CCSRR_{1n} \\ \vdots & \ddots & \vdots \\ CCSRR_{m1} & \cdots & CCSRR_{mn} \end{pmatrix}$$

Now, the last step is the actual retrieval of suitable recommendations for institutional collaboration using strategies 1 and 2. For convenience, Citation weighted T-I matrix, CRR weighted T-I matrix and CCSRR weighted T-I matrix can be also referred to as Citation matrix, CRR matrix and CCSRR matrix.

**Step 6:** *Execution of strategies 1 and 2 for recommendations retrieval*

As discussed in the introduction, the rationale behind the framework is that (i) maximization of strengths in areas of the core competency of the institution through suitable collaborations with institutions that are also having core competency in such areas (*strategy 1*) and (ii) enhancing strengths in areas of the potential core competency of an institution through suitable collaborations with institutions that have (a) core competency in such thematic areas or (b) potential core competency in such thematic areas (*strategy 2*). To achieve this diligently chosen sequence of processes are designed for each of these strategies using Citation, CRR and CCSRR matrices. The algorithm for the execution of strategies 1 and 2 is given below.

1. For an institution $i \in I$, using CRR matrix,
   a) Find the set of strong thematic areas $Q$
      where $Q \subseteq T$ and $q \in Q$ satisfies $CRR_{qi} \geq 1$
   b) Find the set of potential strong thematic areas $U$ using CCSRR matrix
      where, $u' \in U'$ satisfies $CCSRR_{tu'} \geq 1$
      and $U = U' \setminus Q$

### Strategy-I

2. For a thematic area $t \in Q$, using CRR matrix,
   a) Find the set of strong institutions $S$
      where $S \subseteq I$ and $s \in S$ satisfies $CRR_{ts} \geq 1$
3. Initialize $R_i = \emptyset$     **// Recommendations //**
4. $\forall j \neq i \in S$, using Citation matrix,
   if $C_{tj} \geq \delta\, C_{ti}$
   $R_i : R_i \cup j$
   where $\delta = [0.5, 1]$, the threshold for selecting recommendable institutions
   (Preferable value of $\delta$ is 0.75)

### Strategy-II

1. For a thematic area $t \in U$, Using CRR matrix,
   a) Find the set of strong institutions $Y$
      where $Y \subseteq I$ and $y \in Y$ satisfies $CRR_{ty} \geq 1$

   b) Find the set of potential institutions $Z$ using CCSRR matrix
      Where, $z' \in Z'$ satisfies $CCSRR_{tz'} \geq 1$
      and $Z = Z' \setminus Y$
2. Initialize $H_i = \emptyset$     **// High Priority recommendations //**
3. $\forall j \in Y$,
   $H_i : H_i \cup j$
4. Initialize $L_i = \emptyset$     **// Low Priority recommendations //**
5. $\forall j \neq i \in Z$,
   if $CCSRR_{tj} \geq CCSRR_{ti}$
   $L_i : L_i \cup k$

Restricting conditions are imposed for recommendations in the case of strategy 1 using thematic strength and in the case of strategy 2 (low priority recommendations) using CCSRR values to limit the tossing of recommendations to the most feasible ones. When an institution is strong and has a core competency in a particular area, pairing with institutions that are either stronger than it or closer to it in terms of strength (though their strength is somewhat less) will be more beneficial to it. This institution will be recommended to relatively less strong institutions that have core competency or potential core competency in that area and they might extend hands for collaboration. When an institution is relatively less strong or moderately strong in a particular area so that it can be treated to have a potential core competency in that area, pairing with all the institutions that are core competent in that area can be approached for collaboration. Therefore, such recommendations are considered as high priority recommendations in that area and if such collaborations can be achieved,they can be treated as a complementary action. Attempts for collaboration with other institutions having potential core competency also offers some prospects for improvement of the performance of an institution. However, if the thematic area in which the institution is eyeing to improve performance is a relatively well-developed area, there may be too many institutions that have a potential core competency in that area. To limit the institution from being recommended with too many choices, it is offered only with the ones that have higher CCSRR values than itself. If we limit by thematic strength here, some highly reputed institutions that are not so keen to research in this area but received substantially high citations might be tossed to this institution. As such a collaboration is not that so feasible, the usage of CCSRR value for restricting low priority recommendations seems to be sensible.

Now the framework can be demonstrated using a real-world case study of institutions. Before that, details of data collection are discussed.

## 4. Data

Data for each institution is collected from the Web of Science (WoS) database, which is one of the largest online databases that indexes scholarly documents from reputed sources and is thereby regarded as a standard database for bibliometric research. The timewindow used for our analysis is between 2010 and 2019. Computer Science is chosen as the discipline/fieldof analysis and India is chosen as the country of study. Data collection included every type of documents in the database. A filter was imposed in the form of a number of publications. That means institutions that have published at least 25 publications within the period are selected. There were 195 Indian institutions (excluding institution systems like CSIR, IIT systems etc.) that satisfied the criteria making $n=195$. For the use in data processing, metadata fields such as Author Keywords (DE) and Total Times Cited Count (Z9) are included in data collection for each of the distinct publications.

## 5. Results and discussion

As data pre-processing using a multi-layered NLP module is a major step in section 1, it is done prior to the creation of W-K(A) and W-K*(A) networks. NLP module rephrases the author keywords using the word2vec model and thereby reduces the ambiguities to a great extent and converts plurals into their singular counterparts. Now, in the data processing stage, the Network creation and analysis module creates a W-K(A) network by mapping the set of publications W associated with an institution and the set of NLP module processed author keywords K(A). For instance, the W-K(A) network of IIT Kharagpur, one of the pioneer institutions in India, consists of 5307 vertices (1205 works in 1st mode and 4102 keywords in 2nd mode) and 6104 directed arcs or links. W-K* (A) networks are created using Injection methodology and citation information found in the field with tag 'Z9' of the data file of IIT Kharagpur. Thematic strengths of institutions can be computed using the weighted in-degree analysis of the W-K* (A) network, as described in section 2. For instance, the top 20 thematic areas of National Institute of Technology (NIT) Goa and their respective thematic strengths are shown in **table 1**.

Now, one of the crucial steps of the framework is the computation of expertise indices like $x$ and $x(g)$. During this computation process, for each institution, CRRs and CCSRRs will be computed. These will be used for processing in section 2. The computation of the $x$-index of NIT Goa is shown in table 1. It is 17. This means that NIT Goa has got at least 17 thematic areas that are having thematic strengths $\geq 17$. IIT Kharagpur's $x$-index is found to be 116. Thus, 17 and 116 areas can be regarded as the core areas of expertise of NIT Goa and IIT Kharagpur, respectively. $x(g)$-indices of NIT Goa and IIT Kharagpur are 26 and 180, respectively. Computation of $x$ and $x(g)$ index values for NIT Goa is graphically illustrated in fig. 4. Since top $x$ areas are covered in the $x$-core, the thematic areas that fall in the positions $x+1$ to $x(g)$ in the $x(g)$-core can be regarded as the potential areas of the core competency of institutions. Therefore, there are 9 (i.e., 26-17) and 64 (i.e., 180-116) potential core competency areas for NIT Goa and IIT Kharagpur, respectively. The computation of CRRs, CCSRRs, $x$ and $x(g)$ indices mark the end of processes in section 1.

**Table 1. Top 20 thematic areas of research of NIT Goa in terms of thematic strengths**

| S. No. | Keywords (Au) | Thematic strength (Citation counts) | Citation to Rank Ratio (CRRs) |
|---|---|---|---|
| 1 | classifier | 54 | 54 |
| 2 | machine learning | 48 | 24 |
| 3 | sparsity | 45 | 15 |
| 4 | electroencephalogram | 41 | 10.25 |
| 5 | normal and alcoholic eeg signals | 37 | 7.40 |
| 6 | tunable-q wavelet transform | 37 | 6.17 |
| 7 | svm | 37 | 5.29 |
| 8 | object detection | 36 | 4.5 |
| 9 | stroke | 30 | 3.33 |
| 10 | cloud computing | 30 | 3 |
| 11 | optimal ruleset | 28 | 2.54 |
| 12 | diabetes diagnosis | 28 | 2.33 |
| 13 | feature extraction | 25 | 1.92 |
| 14 | wireless sensor network | 25 | 1.79 |
| 15 | ensemble learning | 18 | 1.2 |
| 16 | template matching | 18 | 1.13 |
| **17*** | **accuracy** | **17** | **1** |
| 18 | classification rules | 17 | 0.94 |
| 19 | map estimation | 17 | 0.89 |
| 20 | ultrasound scans | 17 | 0.85 |

**\*$x$-index=17**

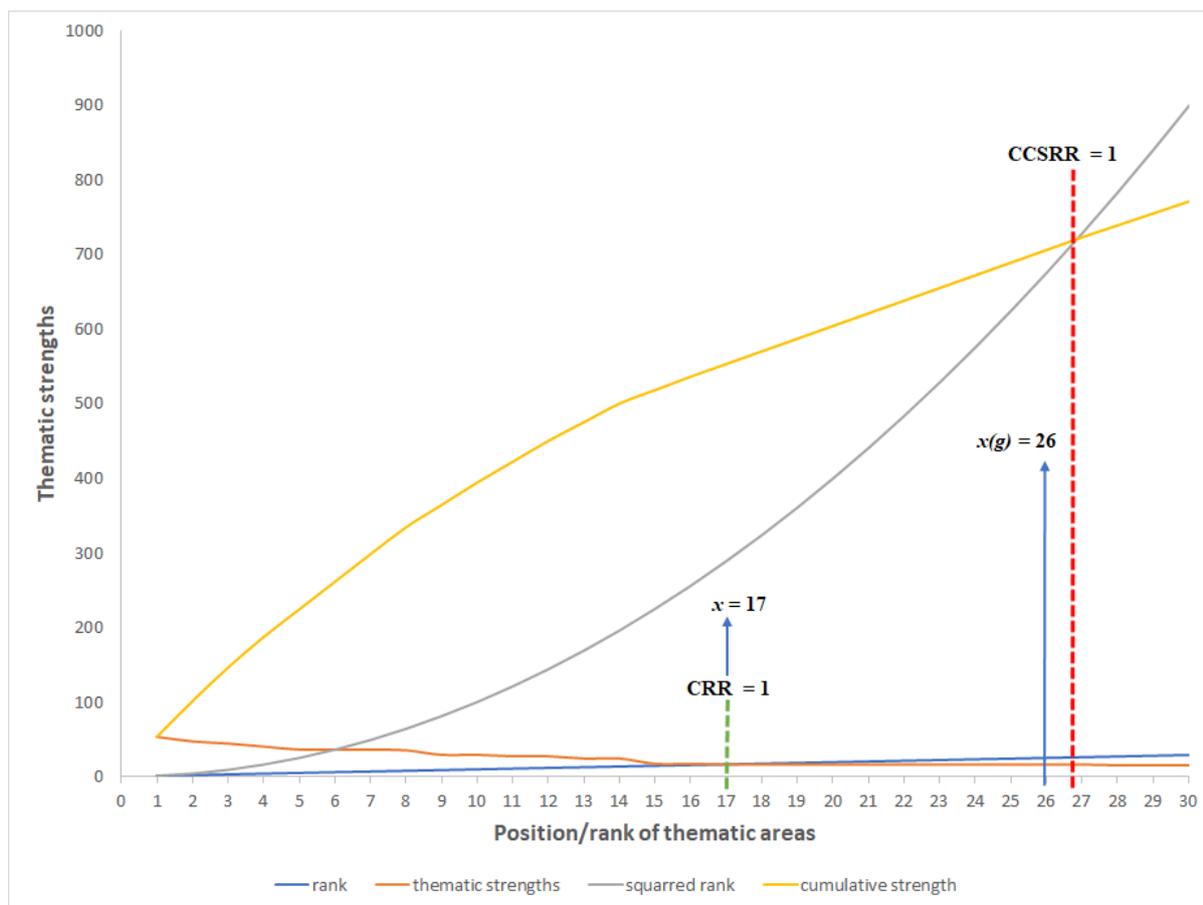

**Figure 4. Graphical illustration of computation of *x* and *x(g)* indices for 'NIT Goa'**

The fourth step, i.e., the determination of a unique number of keywords for the whole set of institutions, is the first process in section 2. The unique number of keywords, m' is found to be 53,750. Using the m' keywords, thematic area institution i.e., K'-I matrix is created. Since there are some uncited keywords, the reduced thematic area-institution matrices or T-I matrices with Citation, CRR and CCSRR weights can be created after the elimination of rows with all zeros. These matrices are of size $46080 \times 195$ each.

Now, for each institution, from its identified core competency areas, thematic areas can be selected and suitable recommendations for expertise maximization (by enhancing its strength in core competency areas) can be retrieved using strategy-1. Then, from its identified potential core competency areas, thematic areas can be selected and suitable high priority recommendations and low priority recommendations (by enhancing its strength in potential core competency areas) can be retrieved using strategy-2. For demonstration, the case of Banaras Hindu University (BHU) is discussed.

From the CRR matrix, the *x*-index of BHU is found to be 41. Therefore, it has 41 core competency areas. From the CCSRR matrix, the *x(g)*-index of BHU is 54 and therefore, BHU have 54-41=13 potential core competency areas. Out of 41 core competency areas, we select the area 'machine learning', one of its core competency areas with 60 citations. Our system is supposed to toss suitable recommendations using strategy-1 from the list of institutions that are strong enough in the area 'machine learning' so that the area falls within their list of core competency areas. Therefore, by step 2 of the algorithm, using the CRR matrix, the set of institutions *S* that have a core competency in thematic area 'machine learning' can be found.

There are 61 institutions in the set $S$. As per step 4, using Citation matrix and taking threshold value $\delta = 0.75$, from $S$, we can select institutions (excluding BHU) that have received citations more than $0.75 \times 60 \approx 45$. Out of 61 eligible institutions, 25 institutions satisfy the criterion and are most suitable for recommendations to enhance the expertise of BHU by improving its core competency. The list of institutions that satisfies this criterion and their thematic strengths in the area 'machine learning' are given in **table 2**.

**Table 2. Recommended institutions to BHU according to strategy-1 for the thematic area 'machine learning'**

| S. No. | Recommendations | Thematic strength |
|---|---|---|
| 1 | VELLORE INSTITUTE OF TECHNOLOGY | 507 |
| 2 | INDIAN INSTITUTE OF TECHNOLOGY IIT DELHI | 398 |
| 3 | INDIAN INSTITUTE OF TECHNOLOGY IIT KHARAGPUR | 317 |
| 4 | AMRITA VISHWA VIDYAPEETHAM | 287 |
| 5 | THAPAR INSTITUTE OF ENGINEERING TECHNOLOGY | 276 |
| 6 | INDIAN STATISTICAL INSTITUTE KOLKATA | 241 |
| 7 | INDIAN INSTITUTE OF TECHNOLOGY IIT MADRAS | 185 |
| 8 | NATIONAL INSTITUTE OF TECHNOLOGY TIRUCHIRAPPALLI | 168 |
| 9 | NATIONAL INSTITUTE OF TECHNOLOGY KURUKSHETRA | 160 |
| 10 | INDIAN INSTITUTE OF TECHNOLOGY IIT GUWAHATI | 129 |
| 11 | INDIAN INSTITUTE OF SCIENCE IISC BANGALORE | 128 |
| 12 | INDIAN INSTITUTE OF INFORMATION TECHNOLOGY ALLAHABAD | 127 |
| 13 | SHANMUGHA ARTS SCIENCE TECHNOLOGY RESEARCH ACADEMY SASTRA | 115 |
| 14 | INDIAN INSTITUTE OF TECHNOLOGY IIT ROORKEE | 113 |
| 15 | INDIAN INSTITUTE OF TECHNOLOGY INDIAN SCHOOL OF MINES DHANBAD | 113 |
| 16 | ANNA UNIVERSITY | 95 |
| 17 | MANIPAL ACADEMY OF HIGHER EDUCATION MAHE | 88 |
| 18 | NATIONAL INSTITUTE OF TECHNOLOGY CALICUT | 87 |
| 19 | BHARATHIAR UNIVERSITY | 86 |
| 20 | GGS INDRAPRASTHA UNIVERSITY | 71 |
| 21 | MALAVIYA NATIONAL INSTITUTE OF TECHNOLOGY JAIPUR | 71 |
| 22 | INDIAN INSTITUTE OF ENGINEERING SCIENCE TECHNOLOGY SHIBPUR IIEST | 67 |
| 23 | KALASALINGAM ACADEMY OF RESEARCH EDUCATION | 58 |
| 24 | JAWAHARLAL NEHRU UNIVERSITY NEW DELHI | 56 |
| 25 | MAULANA AZAD NATIONAL INSTITUTE OF TECHNOLOGY BHOPAL | 49 |

Now, out of the 13 potential core competency areas, we select one area namely 'PSO'. Using the CRR matrix, we have to find the set of strong institutions $Y$ in that area following step 5 a). As per strategy-2 (steps 6 and 7), for high priority recommendations, all institutions in $Y$ can be recommended to BHU. There are 46 such institutions and these are shown in **table 3**. It is interesting to find that BHU's strength in the area 'PSO' is 40 and most of the recommended institutions have strengths greater than BHU in this area.

Institutions that have thematic area 'PSO' as one of their potential core competency areas can be recommended (as low priority recommendations) to BHU if their CCSRR value with respect to the area is greater than or equal to BHU. The set of institutions $Z$ that have 'image retrieval' as a potential core competency area can be found according to step 5 b). Steps 8 and 9 help to retrieve the low priority recommendations. In the case of BHU, since all the institutions in $Z$ are found to have less CCSRR values than BHU with respect to this field, no institutions will be selected for low priority recommendations. This may not be the case for other institutions. For instance, for the institution 'South Asian University', with respect to the field 'fuzzy sets', four institutions namely Indian Institute of Management (IIM) Ahmedabad, Veer Surendra Sai University of Technology, Thapar Institute of Engineering Technology, National Institute of Technology (NIT) Calicut, can be recommended as low priority recommendations. Thus, the working of the framework is demonstrated. Next, we discuss the validation of the framework.

**Table 3. Recommended institutions (high priority) to BHU according to strategy-2 for the thematic area 'PSO (particle swarm optimization)'**

| S. No. | Recommendations | Thematic strength |
|---|---|---|
| 1 | INDIAN INSTITUTE OF TECHNOLOGY IIT DELHI | 509 |
| 2 | INDIAN INSTITUTE OF TECHNOLOGY IIT ROORKEE | 418 |
| 3 | INDIAN INSTITUTE OF SCIENCE IISC BANGALORE | 329 |
| 4 | DELHI TECHNOLOGICAL UNIVERSITY | 269 |
| 5 | INDIAN STATISTICAL INSTITUTE KOLKATA | 263 |
| 6 | INDIAN INSTITUTE OF TECHNOLOGY BHUBANESWAR | 254 |
| 7 | THAPAR INSTITUTE OF ENGINEERING TECHNOLOGY | 228 |
| 8 | UNIVERSITY OF DELHI | 221 |
| 9 | ANNAMALAI UNIVERSITY | 213 |
| 10 | NATIONAL INSTITUTE OF TECHNOLOGY ROURKELA | 201 |
| 11 | NATIONAL INSTITUTE OF TECHNOLOGY DURGAPUR | 189 |
| 12 | ANNA UNIVERSITY | 177 |
| 13 | NATIONAL INSTITUTE OF TECHNOLOGY TIRUCHIRAPPALLI | 176 |
| 14 | THIAGARAJAR COLLEGE OF ENGINEERING | 168 |
| 15 | NETAJI SUBHAS UNIVERSITY OF TECHNOLOGY | 142 |
| 16 | INDIAN INSTITUTE OF TECHNOLOGY IIT KHARAGPUR | 140 |
| 17 | VEER SURENDRA SAI UNIVERSITY OF TECHNOLOGY | 119 |
| 18 | AMITY UNIVERSITY | 113 |
| 19 | UNIVERSITY OF HYDERABAD | 112 |
| 20 | INDIAN INSTITUTE OF TECHNOLOGY IIT KANPUR | 111 |
| 21 | INSTITUTE OF MATHEMATICAL SCIENCES IMSC INDIA | 110 |
| 22 | BIRLA INSTITUTE OF TECHNOLOGY SCIENCE PILANI BITS PILANI | 104 |
| 23 | VELLORE INSTITUTE OF TECHNOLOGY | 104 |
| 24 | SIKSHA O ANUSANDHAN UNIVERSITY | 103 |
| 25 | PERIYAR UNIVERSITY | 98 |
| 26 | KALYANI UNIVERSITY | 94 |
| 27 | PSG COLLEGE TECHNOLOGY | 84 |
| 28 | JAYPEE UNIVERSITY OF INFORMATION TECHNOLOGY | 73 |
| 29 | UNIVERSITY OF BURDWAN | 72 |
| 30 | THE NORTHCAP UNIVERSITY | 69 |
| 31 | SHANMUGHA ARTS SCIENCE TECHNOLOGY RESEARCH ACADEMY SASTRA | 68 |
| 32 | TATA INSTITUTE OF FUNDAMENTAL RESEARCH TIFR | 62 |

| 33 | NATIONAL INSTITUTE OF TECHNOLOGY WARANGAL | 59 |
| --- | --- | --- |
| 34 | SONA COLLEGE OF TECHNOLOGY | 57 |
| 35 | INDIAN INSTITUTE OF TECHNOLOGY IIT PATNA | 56 |
| 36 | MOTILAL NEHRU NATIONAL INSTITUTE OF TECHNOLOGY | 53 |
| 37 | PONDICHERRY UNIVERSITY | 40 |
| 38 | VELAMMAL ENGINEERING COLLEGE | 39 |
| 39 | MEPCO SCHLENK ENGINEERING COLLEGE | 35 |
| 40 | TECHNO INDIA COLLEGE OF TECHNOLOGY | 34 |
| 41 | B S ABDUR RAHMAN CRESCENT INSTITUTE OF SCIENCE TECHNOLOGY | 31 |
| 42 | BANNARI AMMAN INSTITUTE OF TECHNOLOGY | 30 |
| 43 | NATIONAL INSTITUTE OF TECHNOLOGY HAMIRPUR | 30 |
| 44 | SRI KRISHNA COLLEGE OF ENGINEERING TECHNOLOGY | 26 |
| 45 | NORTH EASTERN HILL UNIVERSITY | 21 |
| 46 | RMK ENGINEERING COLLEGE | 10 |

## 5. Evaluation of recommendation system using standard measures

Measures related to desirable attributes of recommendation systems such as *accuracy, novelty, coverage* and *diversity* are identified from related literature as the most used ones for validation of recommendation systems. Precision and Recall (Kent et al., 1955) are two important metrics related to the accuracy in the performance of the recommendation system. In the context of recommendation systems, precision indicates the ratio of retrieved relevant recommendations (true positive) to the total recommendations (true positive + false positive) made. Recall is expressed as the ratio of retrieved relevant recommendations (true positive) to total recommendations that turned out to be relevant which includes the ones that are recommended and possible relevant recommendations that are not recommended (true positive + false negative). This indicates that for measurement of accuracy using both these metrics, information about relevant recommendations is necessary for comparison with the recommendation results to determine true positive, false positive and false negative. In our case, as there is no governing criterion to determine the relevancy of recommendations, evaluation of the accuracy of the framework is not possible at this stage. Even if the information about the materialization of actual collaborations is available after a specific time period, it cannot be used to determine true positive, false positive and negative unless the time period is counted after the implementation of the system and communication of our recommendations to concerned institutions. Thus, accuracy evaluation is beyond the scope of the present work and is reserved post-implementation. However, evaluation of novelty, coverage and diversity of the recommendation system is possible.

For evaluation, we firstly grouped 195 institutions in India into four groups. This grouping is very much essential for diversity evaluation. The *x*-index values of 195 institutions are used for this grouping. Institutions are labelled as 1, 2, 3 and 4, designating the group to which they belong. Grouping is done in the following way:

Group 1: *x*-index $> 80$
Group 2: $80 >$ *x*-index $> 60$
Group 3: $60 >$ *x*-index $> 40$
Group 4: *x*-index $< 40$

195 institutions are distributed among the four groups: 6, 13, 36 and 140 institutions respectively in groups 1, 2, 3 and 4. A validation exercise that maintains the distribution proportion 6: 13: 36: 140 or 1: 2: 6: 23 requires the selection of 32 institutions. It is difficult to show even the relevant results of the analysis of 32 institutions as the recommendations retrieved will also be a large number. So, we are selecting 4 institutions from each group and the validation results related to novelty, coverage and diversity might not be as effective as in the case of an analysis that maintains the proportion 1: 2: 6: 23. However, if the results obtained are sufficient to indicate that the system has enough novelty, coverage and diversity, the validation exercise with the selection we made can be treated as a satisfactory one.

Institutions selected from group 1 are IIT Kharagpur, Thapar Institute of Engineering Technology, IIT Delhi and Vellore Institute of Technology (VIT).

Institutions selected from group 2 are IISc Banglore, IIT Kanpur, NIT Rourkela andIIT Bombay.

Institutions selected from group 3 are NIT Tiruchirapalli, University of Delhi, Jawaharlal Nehru University (JNU) New Delhi and Banaras Hindu University (BHU).

Institutions selected from group 4 arePondicherry University, South Asian University, IIT Hyderabad and IIT Gandhinagar.

*Novelty evaluation*

Evaluation of the ability of systems to toss novel recommendations can be termed as novelty evaluation. In the context of collaborations, information on institutions with which collaboration can be attempted, that cannot be obtained without the help of recommendation algorithms, can be termed as *novel recommendations* (Fouss & Saerens, 2008; Brandão et al., 2013). The novelty of the recommendation system can be evaluated by using the *Novelty index* given by Brandão et al. (2013) in terms of the ability of the system to recommend 'not-so-often' recommended actors. This is based on the assumption that 'often' recommended actors might be known or visible to other actors even without the help of a recommendation system. For the evaluation of novelty of the recommendations made in the case study, we use the novelty index described by Brandão et al. (2013). In the context of institutional collaboration recommendations, the Novelty index is supposed to use the frequency of institutions $f$ in sets of recommended institutions for judging whether the institution is known or not known.

Let $R$ be the unique set of institutions recommended using strategy-1 and $H$ and $L$ respectively be the unique sets of high priority and low priority recommendations retrieved using strategy-2. Now, let the total number of institutions in each unique set be $n_R = |R|$, $n_H = |H|$ and $n_L = |L|$.

Let the unique set of institutions for which recommendations are retrieved using strategy-1 be $CR$ (set of institutions to be communicated about recommendations from the set $R$) and unique sets of institutions for which high priority and low priority recommendations are retrieved using strategy-2 be $CH$ and $CL$. Therefore, the total number of institutions in each unique set will be $n_{CR} = |CR|$, $n_{CH} = |CH|$ and $n_{CL} = |CL|$.

For each member institution $i$ in $R$ or $H$ or $L$, there will be a frequency associated with it (i.e., the number of times it is tossed). Let $f_i$ be that frequency and therefore we can have a set of such of frequencies for $R$, $H$ and $L$ as $F_R$, $F_H$ and $F_L$ respectively.

Novelty index can be defined as the ratio of the median of frequencies of recommendedinstitutions to the size of the target set of institutions (i.e., here the size of the communicated set of institutions).

Let $N_R$, $N_H$ and $N_L$ be the novelty indices with respect to $R$, $H$ and $L$. Then,

$$N_R = \frac{median\ (F_R)}{n_{CR}}$$

Similarly,

$$N_H = \frac{median\ (F_H)}{n_{CH}}$$

$$N_L = \frac{median\ (F_L)}{n_{CL}}$$

Novelty index value lies between [0, 1]. Lower the scores of novelty indices, the higher the novelty in recommendations.

In our case,

$n_R = |R| = 77$, $n_H = |H| = 136$ and $n_L = |L| = 27$

Since we have selected four institutions from four groups $n_{CR}$, $n_{CH}$ and $n_{CL}$ are supposed to be 16. However, if there is no recommendation retrieved for any of the institutions in case of strategy-1 or strategy-2, then that institution will not be present in set $CR$ or $CH$ or $CL$ accordingly and this might reduce the values of $n_{CR}$, $n_{CH}$ and $n_{CL}$. In our case,

$n_{CR} = n_{CH} = 16$ and $n_{CL} = 10$.

Reduction in $n_{CL}$ happened because, for 6 institutions out of the chosen 16, no low priority recommendations were retrieved because for those thematic areas these institutions possessed the top CCSRR values.

As the values of $n_R$ and $n_H$ are relatively high so that the information such as names of institutions in $R$ and $H$ and corresponding frequencies in $F_R$ and $F_H$, the groups in which these institutions belong, etc., cannot be displayed as single tables. For illustrative purposes, we show information related to the set $L$ such as names of institutions and their respective frequencies in $F_L$ in **table 4.**

From table 4, after sorting the frequency set $F_L$ in increasing order, the median of the sorted set $SF_L$ can be computed. It is found to be 1. Now the novelty index with respect to the set $L$ is computed as:

$$N_L = \frac{median\ (SF_L)}{n_{CL}} = \frac{1}{10} = 0.1$$

Similarly median of sets $SF_R$ and $SF_H$ are found to be 2 and 1 respectively. Therefore, novelty indices are:

$$N_R = \frac{median\ (SF_R)}{n_{CR}} = \frac{2}{16} = 0.125$$

$$N_H = \frac{median\ (SF_H)}{n_{CH}} = \frac{1}{16} = 0.0625$$

Since novelty index values for all three sets are found to be very low values (close to zero), recommendations retrieved are substantially novel. This can be indicative of the high novelty of the system.

**Table 4. Set of Low priority recommendations, frequencies of each institution in the set and group identifiers of each institution**

| S. No. | Institution names in the set $L$ | $F_L$ | Group ID. |
|---|---|---|---|
| 1 | NATIONAL INSTITUTE OF TECHNOLOGY WARANGAL | 1 | 4 |
| 2 | INDIAN INSTITUTE OF TECHNOLOGY INDIAN SCHOOLOF MINES DHANBAD | 1 | 2 |
| 3 | ANNA UNIVERSITY | 1 | 2 |
| 4 | NATIONAL INSTITUTE OF TECHNOLOGY PATNA | 1 | 4 |
| 5 | ALIGARH MUSLIM UNIVERSITY | 1 | 4 |
| 6 | INDIAN INSTITUTE OF SCIENCE IISC BANGALORE | 1 | 2 |
| 7 | NATIONAL INSTITUTE OF SCIENCE & TECHNOLOGY (ODISHA) | 1 | 4 |
| 8 | SHRI GURU GOBIND SINGHJI INSTITUTE OF ENGINEERING TECHNOLOGY | 1 | 4 |
| 9 | MALAVIYA NATIONAL INSTITUTE OF TECHNOLOGY JAIPUR | 1 | 3 |
| 10 | NATIONAL ENGINEERING COLLEGE INDIA | 1 | 4 |
| 11 | THAPAR INSTITUTE OF ENGINEERING TECHNOLOGY | 3 | 1 |
| 12 | KALYANI UNIVERSITY | 1 | 4 |
| 13 | VIDYASAGAR UNIVERSITY | 1 | 3 |
| 14 | SARDAR VALLABHBHAI NATIONAL INSTITUTE OF TECHNOLOGY | 1 | 3 |
| 15 | UNIVERSITY OF DELHI | 1 | 3 |
| 16 | TEZPUR UNIVERSITY | 2 | 3 |
| 17 | NATIONAL INSTITUTE OF TECHNOLOGY ROURKELA | 1 | 2 |
| 18 | NATIONAL INSTITUTE OF TECHNOLOGY KARNATAKA | 1 | 4 |
| 19 | UNIVERSITY OF HYDERABAD | 1 | 3 |
| 20 | VISVA BHARATI UNIVERSITY | 1 | 4 |
| 21 | INDIAN INSTITUTE OF TECHNOLOGY IIT BOMBAY | 1 | 2 |
| 22 | VEL TECH RANGARAJAN DR SAGUNTHALA R D INSTITUTEOF SCIENCE AND TECHNOLOGY | 1 | 4 |
| 23 | VEER SURENDRA SAI UNIVERSITY OF TECHNOLOGY | 2 | 3 |
| 24 | SRM INSTITUTE OF SCIENCE TECHNOLOGY | 1 | 4 |
| 25 | CSIR CENTRAL SCIENTIFIC INSTRUMENTS ORGANISATION CSIO | 1 | 4 |
| 26 | INDIAN INSTITUTE OF MANAGEMENT AHMEDABAD | 1 | 4 |
| 27 | NATIONAL INSTITUTE OF TECHNOLOGY CALICUT | 1 | 4 |

*Coverage evaluation*

Coverage evaluation is the evaluation of the ability of the system to toss more distinct recommendations (Gunawardana& Shani, 2015). If the system is able to toss distinct recommendations, the benefits of the recommendation system may not be confined to some of the institutions. Thus, coverage is one of the important desirable characteristics of recommendation systems and there lies the importance of coverage evaluation. Most of the recommendation systems are evaluated for coverage using the set of recommendations only. In cases where systems can select both the set to be recommended and communicated, coverage can be evaluated for both recommended and communicated sets. In cases where systems are able to retrieve more than one set of recommendations, both intra-set and inter-set coverage evaluations can be done. The inventor collaboration recommendation system designed by George et al. (2021) have both these abilities and was evaluated accordingly. In our case, the coverage determination is only required for the set of recommendations as the set of institutions to be communicated is not selected by the system. However, both intra-set and inter-set coverage evaluation can be done for recommendations as our system is able to toss three sets of recommendations. The inter-set and intra-set coverage evaluation can determine whether the recommendations in a set are distinct within itself and with respect to recommendations in another set. *Gini index* (Gunawardana & Shani, 2015), a well-known metric has the ability to reflect the distinction within a set and it can be used as an intra-set coverage indicator. Inter-set coverage can be assessed using the *Jaccard dissimilarity index* and the importance of this evaluation is that if there is no sufficient distinction between the recommended sets, there is a high chance of dominance in occupancy of all the sets by a few institutions. If there is no sufficient distinction within a set of recommendations and between different sets, the recommendations made by the system can turn out to be redundant. Thus, coverage evaluation is in a sense, the investigation of redundancy of the recommendation system.

Gini index (GI) is computed using the proportion of every unique institution in the recommended list and reflects how unequally different the recommendations are to users. Let $P_R(i)$ be the proportion of institution $i$ in the recommendation list $R$, which can be expressed as:

$$P_R(i) = \frac{f_i}{\sum_{i=1}^{n_R} f_i}$$

In our case, $\sum_{i=1}^{n_R} f_i$ is found to be 208, $\sum_{i=1}^{n_H} f_i = 265$ and $\sum_{i=1}^{n_L} f_i = 31$.

The case of $\sum_{i=1}^{n_L} f_i$ can be verified using table 4 where the sum of all the frequencies in $F_L$ adds up to 31.

Let $SP_R$ is the set of sorted (increasing order) proportions of all institutionsin $R$ and if $SP_R(i)$ represent the proportion value at position $i$ of $SP_R$, Gini index of set $R$ can be found out as:

$$GI_R = \frac{1}{n_R} \sum_{i=1}^{n_R} (2i - n_R - 1) \, SP_R(i)$$

Similarly, Gini indices of sets $H$ and $L$ will be:

$$GI_H = \frac{1}{n_H} \sum_{i=1}^{n_H} (2i - n_H - 1) \, SP_H(i)$$

$$GI_L = \frac{1}{n_L} \sum_{i=1}^{n_L} (2i - n_L - 1)\, SP_L(i)$$

Gini index value also lies between [0, 1]. The lower the value of GI, the higher the distinction among recommendations. i.e., there is more equality in the recommendation of all institutions if GI values are low and high value can indicate a high chance for tossing some institutions more often (i.e., redundant recommendations).

Now the GI values of $R$, $H$ and $L$ can be computed as:

$$GI_R = \frac{1}{n_R} \sum_{i=1}^{n_R} (2i - n_R - 1)\, SP_R(i) = \frac{1}{77} \sum_{i=1}^{77} (2i - 77 - 1)\, SP_R(i) = 0.349$$

$$GI_H = \frac{1}{n_H} \sum_{i=1}^{n_H} (2i - n_H - 1)\, SP_H(i) = \frac{1}{125} \sum_{i=1}^{136} (2i - 136 - 1)\, SP_H(i) = 0.327$$

$$GI_L = \frac{1}{n_L} \sum_{i=1}^{n_L} (2i - n_L - 1)\, SP_L(i) = \frac{1}{27} \sum_{i=1}^{27} (2i - 27 - 1)\, SP_L(i) = 0.117$$

From the GI values, the retrieved set of recommendations by strategy-1 is found to be of reasonable coverage but lower than the retrieved sets by strategy-2. Since all these values are found to be less than 0.5, the intra-set coverage of the recommendation system can be regarded as satisfactory. The major reason for relatively low coverage for retrieval by strategy-1 and the high priority retrieval by strategy-2 can be attributed to the emphasis on the recommendation of strong or core competency institutions. If the selection of institutions is based on the proportion 1: 2: 6: 23, GI values might become somewhat lower but the trend ($GI_R > GI_H > GI_L$) might persist. Thus, our intra-set coverage evaluation is indicative of reasonably satisfactory intra-set coverage of the system. Now, we can evaluate the inter-set coverage of the system.

As we have already mentioned, inter-set coverage can be evaluated using the Jaccard dissimilarity index. It is derived from the *Jaccard similarity index* which indicates the similarity between two sets of documents/items. If the presence of common documents/items is high, such sets are more similar and hence high value of similarity can be expected. On the other hand, the Jaccard dissimilarity index takes a high value when there are a greater number of dissimilar or distinct items. Its value lies between [0,1]. When two sets are entirely different, the Jaccard dissimilarity index becomes 1. So, in our case, the closeness of the Jaccard dissimilarity index to 1 indicates sufficient inter-set coverage.

Let $J_D(R,H)$, $J_D(R,L)$ and $J_D(H,L)$ be the Jaccard dissimilarity indices between the sets $R$ and $H$, $R$ and $L$ and $H$ and $L$ respectively. Then,

$$J_D(R,H) = 1 - J(R,H) = 1 - \frac{|R \cap H|}{|R \cup H|} = 1 - \frac{|R \cap H|}{n_R + n_H - |R \cap H|}$$

$$J_D(R,L) = 1 - J(R,L) = 1 - \frac{|R \cap L|}{|R \cup L|} = 1 - \frac{|R \cap L|}{n_R + n_L - |R \cap L|}$$

$$J_D(H,L) = 1 - J(H,L) = 1 - \frac{|H \cap L|}{|H \cup L|} = 1 - \frac{|H \cap L|}{n_H + n_L - |H \cap L|}$$

Where, $J(R,H)$, $J(R,L)$ and $J(H,L)$ are the Jaccard similarity indices between these sets.

In our case, $|R \cap H|$ is found to be 68. $|R \cap L| = 19$ and $|H \cap L| = 22$.

Therefore,

$$J_D(R,H) = 1 - \frac{|R \cap H|}{n_R + n_H - |R \cap H|} = 1 - \frac{68}{77 + 136 - 68} = 1 - \frac{68}{145} = 0.531$$

$$J_D(R,L) = 1 - \frac{|R \cap L|}{n_R + n_L - |R \cap L|} = 1 - \frac{19}{77 + 27 - 19} = 1 - \frac{19}{85} = 0.776$$

$$J_D(H,L) = 1 - \frac{|H \cap L|}{n_H + n_L - |H \cap L|} = 1 - \frac{22}{136 + 27 - 22} = 1 - \frac{22}{141} = 0.844$$

The least dissimilarity is found between $R$ and $H$. This again can be attributed to the emphasis on retrieval of core competency areas for recommendation by strategy-1 and strategy-2 (high priority recommendations). However, as all the dissimilarity values are above 0.5, the inter-set coverage of the system can be considered satisfactory. This might improve if the selection of institutions is based on the proportion 1: 2: 6: 23, however this pattern ($J_D(H,L) > J_D(R,L) > J_D(R,H)$) will remain the same.

On the whole, the coverage of the recommendation system can be treated as sufficient.

*Diversity evaluation*

Diversity evaluation of the recommendation system is the assessment of the ability of the system to toss diverse recommendations, i.e., recommendation of institutions that belong to different classes/groups. Classification of institutions into groups should be rational and pragmatic. As mentioned earlier, our classification is based on the core expertise of institutions and there are four groups labelled as 1, 2, 3 and 4. Diversity evaluation can be done with the help of the famous metric namely, Shannon entropy or Shannon diversity index H (Gunawardana & Shani, 2015). It reflects how well each group/class are represented in the whole set of recommendations. It can be expressed as:

$$H = -\sum_{i=1}^{K} p_i \ln(p_i)$$

Where, $p_i$ is the proportion of class $i$ in the whole set of recommendations and $K$ is the total number of classes.

The higher the value of H or more deviation from 0, the more diverse is the examined system.

The evenness or equitability of the recommendations can be computed using Shannon equitability index E(H). It can be expressed as:

$$E(H) = \frac{H}{\ln(K)}$$

Where $\ln(K)$ is the maximum possible value of H. Therefore, E(H) lies in [0,1]. Closeness of E(H) to 1 indicates more evenness in the distribution of the items among the classes.

In our case, as there are four groups, $K = 4$. The diversity computation for the whole set of low priority recommendations (from which the unique set $L$ is obtained) is illustrated here. From table 4, if we add up the frequencies against one group ID, the total frequency of that group can be computed. The ratio of the total frequency of a group to the sum of total frequencies of all the groups (which is found to be 31) will give the proportion $p_i$ of that group.

Frequencies of groups 1, 2, 3 and 4 are obtained as 3, 5, 9 and 14 and the total number of recommendations in the whole set is 31.

Now, the entropy of the set of low priority recommendations can be computed as:

$$H = - \sum_{i=1}^{K} p_i \ln(p_i) = -(\frac{3}{31} \ln\left(\frac{3}{31}\right) + \frac{5}{31} \ln\left(\frac{5}{31}\right) + \frac{9}{31} \ln\left(\frac{9}{31}\right) + \frac{14}{31} \ln\left(\frac{14}{31}\right)) = 1.238$$

As this value is greater than 0, the set of low priority recommendations is highly diverse.

The equitability index of the set of low priority recommendations is:

$$E(H) = \frac{H}{\ln(K)} = \frac{1.238}{\ln(4)} = 0.893$$

As the Shannon equitability index is found to be close to 1, the distribution of retrieved recommendations is found to be even among the groups.

Similarly, the frequencies of groups 1, 2, 3 and 4 for strategy-1 retrieved recommendations are 53, 49, 67 and 39 respectively and the total number of recommendations is 208. Therefore, its H and E(H) are:

$$H = -(\frac{53}{208} \ln\left(\frac{53}{208}\right) + \frac{49}{208} \ln\left(\frac{49}{208}\right) + \frac{67}{208} \ln\left(\frac{67}{208}\right) + \frac{39}{208} \ln\left(\frac{39}{208}\right)) = 1.368$$

$$E(H) = \frac{H}{\ln(K)} = \frac{1.368}{\ln(4)} = 0.987$$

Strategy-1 retrieved recommendations are found to be of very high diversity and distribution is almost absolutely even among the classes.

In the case of high priority recommendations by strategy-2, the frequencies of groups 1, 2, 3 and 4 are 32, 34, 54 and 112 respectively and the total high priority recommendations are 232. Thus, H and E(H) are:

$$H = -(\frac{37}{265} \ln\left(\frac{37}{265}\right) + \frac{37}{265} \ln\left(\frac{37}{265}\right) + \frac{68}{265} \ln\left(\frac{68}{265}\right) + \frac{123}{265} \ln\left(\frac{123}{265}\right)) = 1.255$$

$$E(H) = \frac{H}{\ln(K)} = \frac{1.255}{\ln(4)} = 0.905$$

High priority recommendations retrieved using strategy-2 are also found to be substantially diverse and evenly distributed among the 4 groups.

As there are no specific rules to properly interpret the scores obtained for evaluation metrics, we specify a rule of thumb. For metrics whose values closer to zero indicate better performance, interpretation can be:

Very High: 0 to 0.15, High: > 0.15 to 0.35, Moderate: > 0.35 to 0.6, Low: > 0.6 to 0.85,

Very Low: > 0.85

For metrics whose values closer to 1 indicates better performance, interpretation can be:

Very Low: 0 to 0.15, Low: > 0.15 to 0.35, Moderate: > 0.35 to High: > 0.6 to 0.85,

Very High > 0.85

The indicative performance of our recommendation system can be summarized as shown in **table 5**. As a whole, our system can be treated as capable of tossing novel and diverse recommendations that are also of satisfactory intra-set and inter-set coverage. The ability of the recommendation system to toss three sets of recommendations ensures the high performance of the system with respect to evaluation metrics except for Inter-set coverage. However, as discussed earlier, from the computed inter-set coverage scores between strategy-1 retrieval, strategy-2 (high priority) retrieval and strategy-2 (low priority) retrieval, the exhibition of moderate inter-set dissimilarity scores between Strategy-1 and Strategy-2 (high priority) recommendations alone will not be a vital indication of compromised performance in terms of coverage.

An interesting observation is that number of scores termed as 'Very High' seems to decline (Strategy-2 (Low Priority) > Strategy-2 (High Priority)> Strategy-1) as the goal of recommendation retrieval turns more ambitious. This can be treated as a reflection of the gradient of difficulty that might occur in the real-world (a kind of indirect issuance of a caution note) to engineer more ambitious collaborations/ties. Thus, an advantage of our system over systems that tosses ambitious recommendations is that our system can toss ambitious recommendations and relatively less ambitious recommendations by appropriately cautioning institutional level decision-makers about the level of difficulty that may incur while proceeding to engineer the recommended collaborations, without compromising the level of performance or maintaining asubstantial level of performance in terms of novelty, coverage and diversity.

**Table 5. Summary of validation of recommendation system**

| Strategy | Performance/Evaluation metric | | | |
|---|---|---|---|---|
| | Novelty | Intra-set coverage | Inter-set coverage* | Diversity & Equitability |
| Strategy-1 | High | High | Moderate, High | Very High |
| Strategy-2 (High priority) | Very High | High | Moderate, High | Very High |
| Strategy-2 (Low priority) | Very High | Very High | High, High | Very High |

*Since there are two inter-set indices for which each strategy is involved, there will be two scores that can be interpreted according to the thumb rule.

## 6. Conclusion

Institutional organization of science and its influence and dependency on other organizations of science such as social organization and intellectual organization of science is vital for the advancement of science. As funding of research happens largely at an institutional level than at any other level, the role and responsibility of institutions to scaleup their contribution in quantity and quality towards scientific progress is steadily increasing. Together with this, the shift witnessed in funding patterns from 'trust-based funding' to 'performance-based funding' forced institutions to devise effective strategies to enhance their performance. Collaboration is known to be one of the effective measures to enhance the performance of partnering institutions. However, the determination of a suitable partner for collaboration is a key factor that determines the success of collaborative ventures. As it is not always an easy task, the importance of recommendation systems lies there. Though scientific literature related to collaboration studies is quite rich and includes many studies related to the development of collaboration recommendation systems for academia using a multitude of approaches including network approach, machine learning, etc., and their combinations, most of these are concentrated on individual/author level collaborations. There are a few studies on co-institutional relationship patterns using networks. But the development of institutional collaboration recommendation systems is almost unexplored. We attempt to bridge this gap by an attempt to design an institutional collaboration recommendation system framework using NLP and a network approach that can be fully developed and implemented as a functional recommendation system.

An institution can have different areas of interest/involvement that spans over many fields. However, its strength or expertise varies from one thematic area to another even within a field. Therefore, the rationale behind our design is that 'the enhancement of an institution can be achieved if it is committed to (i) maximize its strength in its strong or core competency areas and (ii) strive towards the conversion of potential core competency areas to core competency areas by strengthening performance in those areas'. The recommendation algorithm we developed consists of two strategies- Strategy-1 and Strategy-2, that are designed in support of the above-mentioned rationale. The working of the framework is demonstrated successfully and major evaluation metrics related to novelty, coverage and diversity are used to evaluate the performance of the recommendation system. Upon validation, the recommendation system is found to be able to toss recommendations that are (i) excellent in terms of novelty and diversity and (ii) effective in terms of both intra-set and inter-set coverage. Hence, the development of our recommendation system framework as a fully functional recommendation system might be beneficial to a multitude of decision-makers. National level policymakers who plan to develop policies for the enhancement of performance of national institutions in 'thrust-areas' and nurture a consortium/network of centres of excellence by creating a network of expert institutions can make use of our approach. Institutional level policymakers can be direct beneficiaries of our system as it has the potential to enhance institutional performance by tossing suitable recommendations that are of substantial novelty, diversity and coverage. Funding agencies can also benefit from the ability of the system to determine thematic area strengths for 'performance-based funding' as well as 'thrust-area performance' based funding.

Our framework is not free from certain limitations and that serve as an opportunity for further improvement. The multi-layered NLP module has ensured the reduction of ambiguities to a great extent but still, some author keywords that cannot represent thematic areas are successfully evading the NLP processing pipeline. Usage of more advanced NLP techniques when they materialize might improve the performance of the NLP module by ensuring better processing of such keywords. As the framework is capable of utilizing metrics other than citations such as altmetric scores for determining thematic strengths, such endeavours can also be attempted. These are some of the possible improvements of the computational framework that forms the backbone of the full-fledged recommendation system. When it comes to the (software) development and implementation of the system, advanced 'software architecture concepts' such as the one given by Elammari & Elfrjany (2012) that ensures reduced complexity will be considered.

**Acknowledgements**

The authors would like to acknowledge the support provided by the DST-NSTMIS funded project- '*Design of a Computational Framework for Discipline-wise and Thematic Mapping of Research Performance of Indian Higher Education Institutions (HEIs)*', bearing Grant No. DST/NSTMIS/05/04/2019-20, for this work.